\documentclass[aps,prd,nofootinbib,superscriptaddress,preprint,showkeys,showpacs,preprintnumbers]{revtex4}
\usepackage{color}
\usepackage{slashed}
\usepackage{graphics}
\usepackage{graphicx}
\usepackage{dcolumn}
\usepackage{subfigure}
\usepackage{mathrsfs}
\usepackage{bm}
\usepackage{amsmath,amssymb,epsfig}
\usepackage{float}
\allowdisplaybreaks[3]

\newcommand{\bea}{\begin{eqnarray*}}
\newcommand{\eea}{\end{eqnarray*}}
\newcommand{\bean}{\begin{eqnarray}}
\newcommand{\eean}{\end{eqnarray}}

\newcommand{\pa}{\partial}

\usepackage{ulem}
\usepackage[dvipsnames]{xcolor}

\begin{document}

\title{Effective metric of spinless  binaries with radiation-reaction effect  up to fourth Post-Minkowskian order  in effective-one-body theory}

\author{Jiliang {Jing}\footnote{ jljing@hunnu.edu.cn}}
 \affiliation{Department of Physics, Key Laboratory of Low Dimensional Quantum Structures and Quantum Control of Ministry of Education, and Synergetic Innovation
Center for Quantum Effects and Applications, Hunan Normal
University, Changsha, Hunan 410081, P. R. China}
\affiliation{Center for Gravitation and Cosmology, College of Physical Science and Technology, Yangzhou University, Yangzhou 225009, P. R. China}

\author{Weike Deng}
\affiliation{Department
of Physics, Key Laboratory of Low Dimensional Quantum Structures and
Quantum Control of Ministry of Education, and Synergetic Innovation
Center for Quantum Effects and Applications, Hunan Normal
University, Changsha, Hunan 410081, P. R. China}

\author{Sheng Long}
 \affiliation{Department
of Physics, Key Laboratory of Low Dimensional Quantum Structures and
Quantum Control of Ministry of Education, and Synergetic Innovation
Center for Quantum Effects and Applications, Hunan Normal
University, Changsha, Hunan 410081, P. R. China}

\author{Jieci Wang\footnote{ jcwang@hunnu.edu.cn}}
\affiliation{Department
of Physics, Key Laboratory of Low Dimensional Quantum Structures and
Quantum Control of Ministry of Education, and Synergetic Innovation
Center for Quantum Effects and Applications, Hunan Normal
University, Changsha, Hunan 410081, P. R. China}


\begin{abstract}

By means of the scattering angles, we obtain an effective metric of spinless binaries with radiation-reaction effects up to fourth post-Minkowskian order, which is the foundation of the effective-one-body theory. We note that there are freedoms for the parameters of the effective metric because one  equation corresponds to two parameters for each post-Minkowskian order. Accordingly, in order to construct a self-consistent effective-one-body theory in which the  Hamiltonian, radiation-reaction forces and waveforms for the ``plus" and ``cross" modes of the gravitational wave should be based on the same physical model,  we can fix these freedoms by requiring the null tetrad component of the gravitationally perturbed Weyl tensor $\Psi_4^B$ to be decoupled in the effective spacetime.

\end{abstract}


\maketitle

 \section{Introduction}

The gravitational waveform template plays a very important role in the detection of gravitational wave  events \cite{Abbott2016, Abbott2016(2),Abbott2017,Abbott2017(2),Abbott2017(3),Abbott2019,Abbott20211,Abbott20212,Abbott20213,Nitz2021} generated by coalescing binary systems. The basis of the gravitational waveform template is the theoretical model of gravitational radiation, in which the key point is the late dynamical evolution of a coalescing binary system.

In 1999, Buonanno and Damour \cite{Damour1999} introduced a novel approach for studying the gravitational radiation generated from a coalescing compact object binary system by mapping the two-body problem onto an effective-one-body (EOB) problem.
Based on the EOB theory with the  post-Newtonian (PN) approximation, Damour et al. provided an estimate of the gravitational waveforms emitted throughout the inspiral, plunge and coalescence phases~\cite{Damour2000,Damour2000(2)}. The study was then generalized to the case of spinning black holes~\cite{Damour2001, Damour2006}. Later, the EOB waveforms were improved by calibrating the model to progressively more accurate numerical relativity simulations, and spanning larger regions of the parameter space~\cite{Cook,Pan,Pan1,Damour2008,Damour20082, Boyle,Damour20083,Pan2,Pan3,Damour2009,Pan4,Bar,Pan5,Damour2015}, which  plays a vital role in the analysis of the gravitational wave signals \cite{Pan2014,Cao2017,Cai2017}.

To release the assumption  that $v/c$ should be a small quantity in the PN approximation, in 2016, Damour~\cite{Damour2016,Da1} presented another theoretical model by combining EOB theory with the post-Minkowskian (PM) approximation instead of PN approximation. 
Khalil et al.~\cite{Khalil} found that the 4PM dynamics gives better agreement with numerical relativity (NR) than the 3PM dynamics. Damour and Rettegno \cite{Damour2022} compared  NR data on the scattering of equal-mass  binary black holes to analytical prediction based on 4PM dynamics of inspiralling binaries \cite{Dlapa1,2210.05541,Dlapa}, and pointed out that a reformulation of PM information in terms of EOB radial potentials leads to remarkable agreement with NR data, especially when using the radiation-reacted 4PM information.  
 Therefore, this new model may lead to a theoretically improved version of the EOB conservative dynamics, and may be useful in the upcoming era of high signal-to-noise-ratio gravitational-wave observations.

The investigation of the gravitational waveform template using EOB theory consists of two main parts: the dynamical evolution of a coalescing binary system, and the waveforms of the gravitational wave.
The dynamical evolution of a coalescing binary system  for a spinless EOB theory can be described by the Hamilton equation \cite{Damour2000}
\begin{subequations}\label{HEq}
\begin{eqnarray}
&&\frac{dR}{d t} - \frac{\pa H[g_{\mu\nu}^{\text{eff}}]}{\pa P_R}
 = 0,  \\ &&  \frac{d \varphi}{d t} - \frac{\pa H[g_{\mu\nu}^{\text{eff}}]}{\pa P_\varphi}
 = 0,  \\
&& \frac{d P_R}{d t} + \frac{\pa H[g_{\mu\nu}^{\text{eff}}]}{\pa R}
= {\cal F}_R[g_{\mu\nu}^{\text{eff}}], \\ && \frac{d P_\varphi}{d t} = {\cal F}_\varphi[g_{\mu\nu}^{\text{eff}}], 
\end{eqnarray}
\end{subequations}
where  ($R, \varphi$) and ($P_R, P_{\varphi} $) are the polar coordinates and the corresponding  conjugate momenta, respectively, $H[g_{\mu\nu}^{\text{eff}}]$ is the Hamiltonian,  and ${\cal F}_R[g_{\mu\nu}^{\text{eff}}]$ and ${\cal F}_\varphi[g_{\mu\nu}^{\text{eff}}]$ are the  radiation-reaction  forces.  The Hamilton equation \eqref{HEq} shows that, for a  self-consistent EOB theory, the  Hamiltonian, radiation-reaction forces and waveforms  for the ``plus" and ``cross" modes of the gravitational wave should be based on the same effective spacetime.

To get the Hamiltonian $H[g_{\mu\nu}^{\text{eff}}]$,  we should first know the effective metric and the relationship between the energy $\mathcal{E}$ of a real two-body system and that $\mathcal{E}_0$ of a EOB  system.
The radiation-reaction  forces are related to  the rate of the energy loss of the gravitational radiation, which in polar coordinates is described by $\frac{d  E}{d t}= \dot{R}\,{\cal F}_{R} [g_{\mu\nu}^{\text{eff}}]+
\dot{\varphi}\,{\cal F}_{\varphi}[g_{\mu\nu}^{\text{eff}}]$, with
$\frac{dE}{dt}=\frac{1}{4\pi G \omega^2 }\int  |\psi^B_{4}|^2 r^2 d \Omega$ \cite{Ref:poisson,TagoshiSasaki745}.
That is to say, to get the radiation-reaction forces for the ``plus" and ``cross" modes of the gravitational wave, we shall find the energy-loss rate for these modes,
in which the key step is to look for the decoupled equation for the null tetrad component of the gravitationally perturbed Weyl tensor  $\psi^B_{4}=\frac{1}{2}(\ddot{h}_{+}-i\ddot{h}_{\times})$ in the effective spacetime. On the other hand, it is well known that the waveforms for the ``plus" and ``cross" modes of the gravitational wave are also related to $\psi^B_{4}$. These discussions show that we have to know $\psi^B_{4}$ in order to set up a  self-consistent EOB theory.

However, for a general effective spacetime, say the effective metric for the EOB theory based on the PN approximation \cite{Damour2000(2)}, we note that there are three non-vanishing null tetrad components of the trace-free Ricci tensor: $\phi_{00},$ $ \phi_{11} $ and $\phi_{22}$. We cannot get the decoupled equation for   $\psi^B_{4}$ because too many Newman-Penrose quantities are coupled with each other.  Therefore, in  previous studies on waveform template  based on the EOB theory, the Hamiltonian in Eq. \eqref{HEq} is based on the effective metric up to very high PN order, but the radiation reaction forces  are not based on this effective metric.

On the other hand, Damour \cite{Damour} showed that the classical radiation reaction to the emission of gravitational radiation during the large-impact-parameter scattering  of two classical point masses modifies the conservative scattering angle by an additional  radiation-reaction contribution, which  can yield  a finite high-energy 3PM-accurate scattering angle. Dlapa et al. \cite{2210.05541} presented the  4PM  scattering angle with radiation-reaction effect. Therefore, we will also study how the radiation reaction affects the effective spacetime in the following studies.

Recently, we have wanted to do our best to set up a self-consistent EOB theory based on the PM approximation. The foundation of the EOB theory is the effective metric.  In this paper, based on Bern's expression of the conservative Hamiltonian of a relativistic massive spinless two-body system, we  obtain an effective metric with radiation-reaction effects up to the 4PM order. 

The rest of the paper is organized as follows. We present the  scattering angles for the real two-body and EOB systems up to the 4PM order in Sec. II. Then in Sec. III, the mapping relationship between the real relativistic energy $\mathcal{E}$ of the real two-body system and the effective relativistic energy $\mathcal{E}_0$ of the EOB system is investigated, and the effective metric with radiation-reaction effects  up to the 4PM order in  the Schwarzschild-like coordinates is obtained. Finally, conclusions and discussions are presented in the last section.

In this paper,  we take the geometric units with only $c=1$, which is suitable for the calculations in the PM framework.


\section{Scattering angles for real two-body and EOB systems}

The first step to build an EOB theory is to obtain an  effective  metric. In principle, we can find the effective  metric by using one of the action variables, the precession  and the scattering angles.   However, explicit calculations have shown that the scattering angle contains more information than the others, and the parameters of the effective metric are independent of the specific process. Therefore, we will use the scattering angle to find the  effective  metric.

\subsection{Scattering angle with radiation-reaction effect for real two-body system}

The Hamiltonian of a massive spinless binary system   \cite{Da2,Bern20192} is given by
\begin{eqnarray}\label{HR}
H(\vec{p},\vec{r})=\sqrt{|\vec{p}|^{\;2}+m_1^2}
+\sqrt{|\vec{p}|^{\;2}+m_2^2}+\sum_{i=1}^{\infty}c_i
\big{(}\frac{G}{|\vec{r}|}\big{)}^i,
\end{eqnarray}
where $m_1$ and $m_2$  represent the masses of the two particles, $|\vec{p}|^2$ denotes $\vec{p}^{\;2}$,  $\vec{r}$ is the distance vector between particles, and  the explicit expressions of $c_i$ can be found in Refs. \cite{Da2,Bern20192,Dlapa}\footnote{We should note that   $c_i$ in the Ref.  \cite{Da2} is equal to $c_i/i!$ in the Ref. \cite{Dlapa}}. In the following we will use
\begin{align}
&m=m_1+m_2\;,\;\;\;\;\;\; \mu=\frac{m_1m_2}{(m_1+m_2)}\;,\ \ \ \ \ \nu=\frac{m_1m_2}{(m_1+m_2)^2}\;,\nonumber\\
&E_1=\sqrt{\vec{p}^{\;2}+m_1^2}\;,\
E_2=\sqrt{\vec{p}^{\;2}+m_2^2}\;, \ \ \  \; E=E_1+E_2.
\end{align}
In spherical coordinates $\{t, r,\theta,\phi\}$, we have
$\vec{p}^{\;2}=p_r^2+\frac{p_{\theta}^2}{r^2}+
\frac{p_{\phi}^2}{r^2\sin^2\theta},$  $
\vec{r}\cdot\vec{p}=r p_r,$ and  $
\vec{p}^{\;2}\vec{r}^{\;2}-(\vec{r}\cdot\vec{p})^2
=p^2_{\theta}+\frac{p_{\phi}^2}{\sin^2\theta}.$

There are two conservative quantities, the energy $\mathcal{E}=-p_t$ and the angular momentum $J=p_{\phi}$ of the real two-body system, associated with the  Hamiltonian (\ref{HR}).  Without loss of generality, we set $\theta=\frac{\pi}{2}$ to denote the plane in which two particles reside.
 Then, we can express the reduced action as
\begin{align}
 S=-\mathcal{E}t+J\phi+S_r(r,\mathcal{E},J)\;.
\end{align}
Using the Hamilton-Jacobi equation
$
\frac{\partial S}{\partial t}+H(q,\frac{\partial S}{\partial q})=0
$,
we can obtain the radial momentum $p_r^2=(\frac{dS_r}{dr})^2$ as a function of the radial coordinate $r$, which up to the 4PM order is given by
\begin{eqnarray}\label{prsquare}
p_r^2&=&\frac{P_0r^2-J^2}{r^2}+P_1\Big(\frac{G}{r}\Big) +P_2\Big(\frac{G}{r}\Big)^2+P_3\Big(\frac{G}{r}\Big)^3+P_4\Big(\frac{G}{r}\Big)^4\;,
\end{eqnarray}
with
\begin{eqnarray}
P_0&=&\Big(\frac{
\mu }{ \Gamma}\Big)^2(\gamma^2-1)\;,\label{p0}\nonumber \\
P_{1}
&=&2 m \mu^2 \Big(\frac{2 \gamma^2-1}{\Gamma}\Big)\;,\label{p1}\nonumber\\
P_2&=&\frac{3 m^2 \mu^2}{2}\Big(\frac{5 \gamma^2-1}{\Gamma}\Big)\;,\label{p2}\nonumber\\
P_3&=&\Gamma m^3 \mu^2 P_{30},
\label{p3} \nonumber \\
P_4 &=& 2 m_1^4 m_2^4 f_4 \,,
   \label{P4}
\end{eqnarray}
where   $\gamma = \frac{1}{2} \frac{\mathcal{E}^2-m_1^2-m_2^2}{m_1m_2} $,  $   \Gamma =  \frac{\mathcal{E}}{m_1+m_2}$,  $P_{30} = \frac{18\gamma^2-1}{2\,\Gamma^2}+\frac{8\,\nu\, (3+12\gamma^2-4\gamma^4)}{\Gamma^2\, \sqrt{\gamma^2-1}} \mbox{arcsinh}\sqrt{\frac{\gamma-1}{2}}+\frac{\nu}{\Gamma^2}\big(1-\frac{103}{3}\gamma-48 \gamma^2-\frac{2}{3} \gamma^3+\frac{3 \,\Gamma\,(1-2\gamma^2)(1-5 \gamma^2)}{(1+\Gamma)(1+\gamma)}\big)$,  and the explicit form of $f_4$ (which is too long, so we do not present it here) can be found in the ancillary file of Ref. \cite{2210.05541}.

Then, by using the  definition 
\begin{align}
\chi^{\text{Nor}}=-\pi+2 J \int_{r_{\rm min}}^{\infty}\frac{d r}{r^2\sqrt{p_r^2}}\;, \label{chireal}
\end{align}
where the minimum distance $r_\text{min}$ is determined by $p_r(r_{min})=0$, we can obtain the scattering angle up to 4PM order without the radiation-reaction effect, which  is  described as
\begin{eqnarray}
\chi^{\text{Nor}}=\chi^{\text{Nor}}_1\frac{G}{J}+\chi^{\text{Nor}}_2\Big(\frac{G}{J}\Big)^2+\chi^{\text{Nor}}_3 \Big(\frac{G}{J}\Big)^3+\chi^{\text{Nor}}_4 \Big(\frac{G}{J}\Big)^4  \;,
\end{eqnarray}
with
\begin{eqnarray}
&&\chi^{\text{Nor}}_1 = 2 \,m_1\, m_2 \,\frac{2\gamma^2-1}{\sqrt{\gamma^2-1}}, \  \nonumber \\  &&\chi^{\text{Nor}}_2=\frac{3\,\pi \,m_1^2\, m_2^2}{4}\, \frac{5\gamma^2-1}{\Gamma},\ \nonumber \\
&&\chi^{\text{Nor}}_3= 2\, m_1^3\, m_2^3\, \sqrt{\gamma^2-1}\, P_{30} +\frac{2}{\pi}\chi^{\text{Nor}}_1 \, \chi^{\text{Nor}}_2-\frac{(\chi^{\text{Nor}}_1)^3}{12} , \nonumber \\
&& \chi^{\text{Nor}}_4=\frac{3 \pi}{4}m_1^4 m_2^4 f_4+\frac{3\pi}{8}\chi^{\text{Nor}}_1 \chi^{\text{Nor}}_3 +\frac{3}{2\pi}(\chi^{\text{Nor}}_2)^2-\frac{3}{4}(\chi^{\text{Nor}}_1)^2 \chi^{\text{Nor}}_2+\frac{\pi}{32}(\chi^{\text{Nor}}_1)^4. \label{2pmchi}
\end{eqnarray}

On the other hand, Damour \cite{Damour}  presented the  following 3PM  radiation-reaction effect
\begin{eqnarray}
  \chi_3^{rr}&=& -\frac{2 \nu }{\Gamma^2}\frac{\gamma(2\gamma^2-1)^{3/2}}{3 (\gamma^2-1)^2}\Big[\frac{(5\gamma^2-8)\sqrt{\gamma^2-1}}{\gamma}+2 (9-6 \gamma^2) \mbox{arcsinh}\sqrt{\frac{\gamma-1}{2}}\Big], \end{eqnarray}
and  Dlapa et al. \cite{2210.05541} obtained the 4PM  radiation-reaction effect $\chi^{\text{rr}}_4$, which is too long, so we do not list it here. The explicit form of  $\chi^{\text{rr}}_4$ can be found in the ancillary file of Ref. \cite{2210.05541}.

It is worth noting that the relation between our results  $\chi^{\text{Nor}}_i$ and the  $\chi _j^{(i)}$  of Refs.  \cite{Kalin,Kalin1, Dlapa} is given by  $\chi^{\text{Nor}}_i=2 (m_1 m_2)^i \chi _j^{(i)}$, since the PM coefficients of the scattering angle are defined by $\frac{\chi}{2}=\sum_{i=1} \chi_j^{(i)}/j^i$ with $j=J/(G m_1 m_2 )$ in Refs. \cite{Kalin,Kalin1, Dlapa} (we do not replace $J$ with $j$ because we will find the relation between  $J$ and  $J_0$, where  $J_0$ is the angular momentum  of the effective system). Accordingly, the coefficients of the total  scattering angle with radiation-reaction effects  up to 4PM order for the real two-body system are
\begin{eqnarray}
\chi^{\text{rel}}_1 &=& 2 \,m_1\, m_2 \,\frac{2\gamma^2-1}{\sqrt{\gamma^2-1}}, \  \nonumber \\  \chi^{\text{rel}}_2&=&\frac{3\,\pi \,m_1^2\, m_2^2}{4}\, \frac{5\gamma^2-1}{\Gamma},\ \nonumber \\
\chi^{\text{rel}}_3&=& 2\, m_1^3\, m_2^3\,\big(\sqrt{\gamma^2-1}\, P_{30}+\chi_3^{rr}\big) +\frac{2}{\pi}\chi^{\text{rel}}_1 \, \chi^{\text{rel}}_2-\frac{(\chi^{\text{rel}}_1)^3}{12} , \nonumber \\
 \chi^{\text{rel}}_4&=&2 m_1^4 m_2^4\chi^{\text{rr}}_4-\frac{3\pi }{4}m_1^3 m_2^3\chi^{\text{rel}}_1 \chi^{\text{rr}}_3+\frac{3 \pi}{4}m_1^4 m_2^4 f_4+\frac{3\pi}{8}\chi^{\text{rel}}_1 \chi^{\text{rel}}_3 +\frac{3}{2\pi}(\chi^{\text{rel}}_2)^2\nonumber \\    &-&\frac{3}{4}(\chi^{\text{rel}}_1)^2 \chi^{\text{rel}}_2+\frac{\pi}{32}(\chi^{\text{rel}}_1)^4.
\label{2pmchi1}
\end{eqnarray}

\subsection{Scattering angle  for  EOB system  in Schwarzschild-like coordinates}

The scattering angle $\chi^{\text{eff}}$ for an EOB system can be found from the dynamics of a test particle scattered by  a black hole described by  the effective metric $g_{\mu\nu}^{\rm eff}$  which can be expressed as
\begin{eqnarray}
ds_{\rm eff}^2=A dt^2-\frac{D^2}{A}dR^2- R^2(d\theta^2+\sin^2\theta d\phi^2),\label{Mmetric}
\end{eqnarray}
with
\begin{eqnarray}\label{expansionB}
&& A=1+\sum_{i=1}^\infty a_i \Big(\frac{GM_0}{R}\Big)^i, \nonumber \\ && D=1+\sum_{i=1}^\infty d_i \Big(\frac{GM_0}{R}\Big)^i,
\end{eqnarray}
where  $M_0$ is the mass of the black hole, and $a_i$ and $d_i$ ($i=1, 2, 3, 4$) are dimensionless parameters which will be found in the following equations.

In the effective spacetime, the effective Hamilton-Jacobi equation reads
\begin{eqnarray}
g^{\mu\nu}_{\rm eff}\frac{\partial S_{\rm eff}}{\partial x^{\mu}}
\frac{\partial S_{\rm eff}}{\partial x^{\nu}}+m_0^2=0\;.\label{374H}
\end{eqnarray}
Since the effective spacetime described by the line element  (\ref{Mmetric}) has spherical symmetry,  without loss of generality, in the equatorial plane ($\theta=\frac{\pi}{2}$), $S_{\rm eff}$ reduces to
\begin{eqnarray}
S_{\rm eff}=-\mathcal{E}_0t+J_0\Phi+S_R^{0}(R,\mathcal{E}_0,J_0)\;,\label{375H}
\end{eqnarray}
where $\mathcal{E}_0$ and $J_0$ are the effective energy and angular momentum, respectively.
Substituting Eq.~\eqref{375H} into Eq.~\eqref{374H}, we obtain
\begin{eqnarray}
-\frac{\mathcal{E}_0^2}{A}+\frac{A}{D^2}
\Big{(}\frac{dS_R^0(R,\mathcal{E}_0,J_0)}{dR}\Big{)}^2+
\frac{J_0^2}{R^2}+m_0^2=0\;,
\end{eqnarray}
and which leads to
\begin{eqnarray}
\frac{dS_R^0(R,\mathcal{E}_0,J_0)}{dR}=\sqrt{\mathcal{R}_0(R,\mathcal{E}_0,J_0)}\;,
\end{eqnarray}
with
\begin{eqnarray}
\mathcal{R}_0(R,\mathcal{E}_0,J_0)=
\frac{D^2}{A^2}\mathcal{E}_0^2-\frac{D^2}{A}
\bigg{[}m_0^2+\frac{J_0^2}{R^2}\bigg{]}\;.\label{effR0}
\end{eqnarray}
Substituting Eq.~\eqref{expansionB} into Eq.~\eqref{effR0}, up to the 4PM order, we find $\mathcal{R}_0(R,\mathcal{E}_0,J_0)$ can be expressed as
\begin{eqnarray}\label{eobaction}
\mathcal{R}_0(R,\mathcal{E}_0,J_0)&=&R_{00}-\frac{J_0^2}{R^2}+\frac{R_{11} G}{R}+\frac{R_{22} G^2}{R^2}-\frac{R_{31} J_0^2 G -R_{33} G^3 }{R^3}\nonumber \\&-&\frac{R_{42}J_0^2 G^2-R_{44} G^4}{R^4}-\frac{R_{53}J_0^2 G^3}{R^5}-\frac{R_{64}J_0^2 G^4}{R^6}\;,
\end{eqnarray}
with
\begin{eqnarray}\label{Rii}
R_{00}&=&\mathcal{E}_0^2-m_0^2\;,\nonumber \\
R_{11}&=&2 M_0\left[2 \mathcal{E}_0^2-m_0^2\right] \;,
\nonumber \\
R_{22}&=&M_0^2\left[(a_2-2(2+d_2)) m_0 ^2+2 \mathcal{E}_0^2\left(6-a_2+d_2\right)\right]\;,\nonumber \\
R_{31}&=&2 M_0  \:,\nonumber\\
R_{33}&=&M_0^3\left[(4 a_2+a_3-2(4+2 d_2+d_3)) m_0^2+2 \mathcal{E}_0^2\left(16-6a_2-a_3+4 d_2+d_3\right)\right]\;,\nonumber \\
R_{42}&=&(4-a_2+2 d_2 )  M_0^2\;,\nonumber \\
R_{44}&=&
M_0^4 \Big\{-(16-12 a_2+a_2^2-4 a_3 -a_4 +8 d_2-2a_2 d_2+d_2^2+4 d_3+ 2 d_4) m_0^2\nonumber \\ &+& \mathcal{E}_0^2  \Big[80+ 3 a_2^2-12 a_3-2 a_4 +24 d_2+d_2^2-4 a_2 (12+d_2)
 +8 d_3+2 d_4\Big]\Big\}\;,\nonumber \\
R_{53}&=&(8-4 a_2-a_3+4 d_2+2 d_3) M_0^3\;,\nonumber \\
R_{64}&=&(16-12 a_2+a_2^2-4 a_3 -a_4 +8 d_2-2a_2 d_2+d_2^2+4 d_3+ 2 d_4)  M_0^4\;,
\end{eqnarray}
where  we have taken $a_1=-2$ and $d_1=0$ by requiring that the effective metric at the 1PM order coincides with the Schwarzschild metric for the sake of concise expressions.

For the EOB system, the  scattering angle can be expressed as
\begin{eqnarray}\label{chii}
\chi^{\text{eff}}&=&-\pi-2  \int_{R_{\rm min}}^{\infty}\frac{\partial\sqrt{ \mathcal{R}_s (R,\mathcal{E}_0,J_0)}}{\partial J_0} d R.
\end{eqnarray}
where $R_\text{min}$ is the minimum distance  determined by setting the vanishing of $\mathcal{R}_s (R,\mathcal{E}_0,J_0)$.
Substituting Eq. (\ref{eobaction})  into Eq. (\ref{chii})  and  working out the integration up to the 4PM order, we get
\begin{eqnarray}\label{chiii}
\chi^{\text{eff}}&=&\frac{R_{11}}{\sqrt{R_{00}}}\frac{G}{J_0}+
\frac{ \pi \big[2 R_{22} -  R_{11} R_{31} + R_{00}(\frac{3}{4}  R_{31}^2-R_{42})\big]}{4}
 \Big(\frac{G}{J_0}\Big)^2\nonumber \\
  &+&\frac{1}{12 R_{00}^{3/2}} \Big\{-R_{11}^3 - 6 R_{00} R_{11}^2 R_{31} + 12 R_{00} R_{11} \Big[R_{22} \nonumber \\ &+& 2 R_{00} (R_{31}^2 - R_{42})\Big ] -
   8 R_{00}^2 \Big[3 R_{22} R_{31} - 3 R_{33} \nonumber \\
 &+ &2 R_{00} (R_{31}^3 - 2 R_{31} R_{42} + R_{53})\Big]\Big\}\Big(\frac{G}{J_0}\Big)^3\nonumber\\
&+& \frac{3\pi }{1024}\Big\{128 R_{22}^2+48 R_{11}^2 \left(5 R_{31}^2-4 R_{42}\right)
\nonumber\\
&-&96 R_{22} \left(4 R_{11} R_{31}-5 R_{00}
R_{31}^2+4 R_{00} R_{42}\right)
\nonumber\\
&+&16 R_{11} \left(16 R_{33}+R_{00}\left(-35 R_{31}^3+60 R_{31} R_{42}-24
R_{53}\right)\right)
\nonumber\\
&+&R_{00} \Big[-384 R_{31} R_{33}+256 R_{44}+3 R_{00} \Big(105 R_{31}^4
\nonumber\\
&-&280 R_{31}^2
R_{42}+80 R_{42}^2+160 R_{31} R_{53}-64 R_{64}\Big)\Big]\Big\} \Big(\frac{G}{J_0}\Big)^4.
 \end{eqnarray}
Then, using Eqs. (\ref{Rii}) and (\ref{chiii}) we have
\begin{eqnarray}\label{x} \chi^{\text{eff}}=\chi^{\text{eff}}_1\frac{G}{J_0}+\chi^{\text{eff}}_2 \Big(\frac{G}{J_0}\Big)^2+\chi^{\text{eff}}_3 \Big(\frac{G}{J_0}\Big)^3+\chi^{\text{eff}}_4 \Big(\frac{G}{J_0}\Big)^4,
 \end{eqnarray}
 with
\begin{align}\label{2pmchii}
\chi^{\text{eff}}_1&=\frac{2 M_0 \left[2 \mathcal{E}_0^2-m_0^2 \right]}{\sqrt{\mathcal{E}_0^2-m_0^2}},  \nonumber \\
\chi^{\text{eff}}_2&=\frac{M_0^2 \pi}{4}\bigg{[}\mathcal{E}_0^2(15 - 3 a_2 + 2 d_2)+m_0^2( a_2-3 - 2 d_2)\bigg{]}, \nonumber  \\
\chi^{\text{eff}}_3&=\frac{2\chi^{\text{\text{\text{eff}}}}_1 \chi^{\text{eff}}_2}{\pi}-\frac{\chi^{\text{eff\,3}}_1}{12} -\frac{M_0^3 \sqrt{\mathcal{E}_0^2-m_0^2} }{3}
[(3 - 3 a_2 - 2 a_3 \nonumber \\ &   - 2 d_2 + 4 d_3)     m_0^2 + 2 \mathcal{E}_0^2( 15 a_2 + 2(2 a_3 -  d_2 -  d_3) -27)],  \nonumber\\
\chi^{\text{eff}}_4&=\frac{3\pi}{8}\chi^{\text{eff}}_1 \chi^{\text{eff}}_3 +\frac{3}{2\pi}(\chi^{\text{eff}}_2)^2-\frac{3}{4}(\chi^{\text{eff}}_1)^2 \chi^{\text{eff}}_2+\frac{\pi}{32}(\chi^{\text{eff}}_1)^4 + \frac{\pi M_0^4 (\mathcal{E}_0^2-m_0^2)}{64}\Big[(387-390 a_2  \nonumber \\
&   +51 a_2^2  -164 a_3-60 a_4+12 a_2 d_2-4 d_2(17+6 d_2)-8 d_3+24 d_4) \mathcal{E}_0^2-(3+3(a_2-2)a_2\nonumber \\ & - 4 a_3-12 a_4+12 a_2 d_2-4 d_2 (1+6 d_2)-40 d_3+24 d_4)m_0^2 \Big]. \end{align}

\section{Energy map and effective metric with radiation-reaction for  EOB theory}

In the EOB theory \cite{Damour1999}, the main idea is to map the two-body problem onto an EOB problem, i.e.,  a test particle orbits around a massive black hole described by effective metric. This map can be realized by identifying the scattering angles for the two systems order by order, i.e., by taking $\chi^{\text{real}}_i=\chi^{\text{eff}}_i\;, $ for $i=1,~2,~3,~4,\cdot \cdot\cdot $.  Now, we will find the energy map and the effective metric with radiation-reaction effects for  EOB theory.

\subsection{Energy map for  EOB theory}

Following Buonanno and Damour \cite{Damour1999,Damour2016}, we take
\begin{eqnarray}
&&m_0=\frac{m_1m_2}{m_1+m_2},\label{i1}\nonumber \\
&&M_0=m_1+m_2,\nonumber \\
&&J_0=J.\label{i2}
\end{eqnarray}
Then, by requiring that the effective metric at the 1PM order coincides with the Schwarzschild metric, and  taking
\begin{eqnarray}
\chi^{\text{real}}_1=\chi^{\text{eff}}_1\;,\label{chi1}
\end{eqnarray}
we find
\begin{eqnarray}
\mathcal{E}_0&=\frac{\mathcal{E}^2-m_1^2-m_2^2}{2(m_1+m_2)}\label{energymap1}\;, \end{eqnarray}
which is the energy map between the relativistic energy $\mathcal{E}$ of the real two-body system and the relativistic energy $\mathcal{E}_0$ of the EOB system.

By taking the reduced energies as $\hat{\mathcal{E}_0}=\mathcal{E}_0/m_0$ and $\hat{\mathcal{E}}=\mathcal{E}/m_0$, Eq. (\ref{energymap1}) can be rewritten as
\begin{eqnarray}
\hat{\mathcal{E}_0}=\frac{\nu}{2}\hat{\mathcal{E}}^2-\frac{1}{2 \nu}+1, \end{eqnarray}
which is useful in the related calculations.

\subsection{Effective metric with radiation-reaction effect in  EOB theory}

We note that there are freedoms for the parameters  of the effective metric because one  equation corresponds to two parameters ($a_i$, $d_i$) for each post-Minkowskian order $i$. The freedoms can be fixed in the following discussions.

Substituting Eqs. \eqref{2pmchi1} and \eqref{2pmchii} into 
\begin{eqnarray}
\chi^{\text{real}}_i=\chi^{\text{eff}}_i\;, \ \ \ \ \text{(i=2, 3, 4)},\label{chi2}
\end{eqnarray}
we find that the parameters $a_i$ in the effective metric are given by
\begin{eqnarray} \label{parameters}
a_2&=&\frac{3(1- \, \Gamma)(1-5 \, \gamma^2)}{\Gamma\, (3\, \gamma^2-1 )}\, ,\nonumber \\
a_3&=&\frac{3}{2 (4 \gamma^2-1 )}\Big[\frac{3- 2\, \Gamma-3 (15 -8\, \Gamma ) \gamma^2+6 ( 25-16 \, \Gamma ) \gamma^4}{\Gamma\, (3\, \gamma^2-1 )}-2 P_{30}-\frac{2 \chi_3^{rr}}{\sqrt{\gamma^2-1}}\Big]\nonumber \\ &&-d_2 \frac{4-34 \gamma^2+24 \gamma^4}{1-7  \gamma^2+12 \gamma^4 }+d_3\frac{ 2 (1-  \gamma^2)}{1-4 \gamma^2}, \nonumber \\
a_4&=&\frac{1}{(\gamma^2-1)(5 \, \gamma^2-1)}\Big\{-4 f_4-\frac{32 \chi_4^{rr}}{3 \pi}+\frac{8(2 \gamma^2-1) \chi_3^{rr}}{\sqrt{\gamma^2-1}} +\frac{1}{12}\Big[ (\gamma^2-1)\Big( -3+387 \gamma^2\nonumber \\  &&
  +6 a_2 (1-65  \gamma^2) +3 a_2^2 (17  \gamma^2-1)+4 a_3 (1-41  \gamma^2)\Big) \Big]\Big\} +\frac{1}{3(5\gamma^2-1)}\Big[d_2-3 a_2 d_2 \nonumber \\ &&+6 d_2^2+10d_3-6 d_4-((17-3 a_2 +6 d_2 )d_2 +2 d_3 -6 d_4 )\gamma^2 \Big],     \end{eqnarray}
where the parameters $a_3$ and $a_4$ include the  terms $\chi^{rr}_3$ and $\chi^{rr}_4$, which  represent the 3PM and 4PM radiation-reaction effects, respectively. That is to say,  the radiation reaction will affect the structure of the effective spacetime.

How to fix $d_i$? Before fixing this issue, let us determine how we can fix the parameters of the effective metric in the EOB theory based on the PN approximation. It was shown that the number of equations is less than the number of parameters of the effective metric at the 3PN order, then some parameters were chosen artificially  \cite{Damour2000(2)}.  It is well known that the binding energy and the gravitational wave energy flux are two central ingredients that enter the computation of gravitational waveforms.  Therefore,  we have thought that the parameters  $d_i$ can be fixed by comparing the EOB  predictions for the binding energy of a two-body system on a quasicircular inspiraling orbit against the results of numerical relativity simulations. However, we have found that, along this way, we will lose the self-consistency for an EOB theory, for the following reasons:
for the general effective metric \eqref{Mmetric} with the parameters \eqref{parameters} there are three  non-vanishing null tetrad components of the tracefree Ricci tensor: $\phi_{00},$ $ \phi_{11} $ and $\phi_{22}$.
We cannot obtain a decoupled equation for   $\psi^B_{4}$ in the effective spacetime because too many Newman-Penrose quantities are coupled with each other.   That is to say, we can use the effective metric to construct the Hamiltonian, but we cannot obtain the radiation reaction forces  and the waveforms for the ``plus" and ``cross" modes based on the effective metric.

However, if we take $ D=1$ (i.e., $d_i=0$ for $i=2,3,4$) which fixed the freedoms of the parameters of the effective metric, we have
\begin{eqnarray} \label{parameters1}
a_2&=&\frac{3(1- \, \Gamma)(1-5 \, \gamma^2)}{\Gamma\, (3\, \gamma^2-1 )}\, ,\nonumber \\
a_3&=&\frac{3}{2 (4 \gamma^2-1 )}\Big[\frac{3- 2\, \Gamma-3 (15 -8\, \Gamma ) \gamma^2+6 ( 25-16 \, \Gamma ) \gamma^4}{\Gamma\, (3\, \gamma^2-1 )}-2 P_{30}-\frac{2 \chi_3^{rr}}{\sqrt{\gamma^2-1}}\Big], \nonumber \\
a_4&=&\frac{1}{(\gamma^2-1)(5 \, \gamma^2-1)}\Big\{-4 f_4-\frac{32 \chi_4^{rr}}{3 \pi}+\frac{8(2 \gamma^2-1) \chi_3^{rr}}{\sqrt{\gamma^2-1}} +\frac{1}{12}\Big[ (\gamma^2-1)\Big( -3+387 \gamma^2\nonumber \\  &&
  +6 a_2 (1-65  \gamma^2) +3 a_2^2 (17  \gamma^2-1)+4 a_3 (1-41  \gamma^2)\Big) \Big]\Big\}.     \end{eqnarray}
Then we have $\phi_{00}=\phi_{22}=0 $, and we can obtain the decoupled equation for the null tetrad component of the gravitationally perturbed Weyl tensor $\Psi_4^B$ by means of the Refs. \cite{Jing,Jing1,Jing2}, in which we have shown that the decoupled equation for $\Psi_4^B$ can be obtained if the metric takes the form $B=D^2/A=1/A$ up to any PM order. In this way, we can set up  a  self-consistent EOB theory in which the Hamiltonian, radiation-reaction forces  and waveforms for the ``plus" and ``cross" modes are based on the same effective spacetime.

\section{Conclusions and discussions}\label{sec5}

With the help of the scattering angles, we obtained the effective metric  for spinless  binaries with radiation-reaction effects up to the 4PM order in the EOB  theory. We can show that these results are consistent by means of  the action variables, the precession and scattering angles, which implies that parameters of the effective metric are independent of any specific process.

Specifically, we can summarize our results as follows.
The relations for the masses and angular momenta between the EOB and the real two-body systems are  taken as
$ m_0=\frac{m_1m_2}{m_1+m_2}, $ $   M_0=m_1+m_2,$  and $
J_0=J.$ Then,
by requiring that the effective metric at the 1PM order coincide with the Schwarzschild metric, and taking $\chi^{\text{real}}_1=\chi^{\text{eff}}_1$, we find that the mapping relationship between the relativistic energies of the real two-body and the EOB systems is described by
$
\mathcal{E}_0=\frac{\mathcal{E}^2-m_1^2-m_2^2}{2(m_1+m_2)},
$
which  holds for the 4PM order and we conjecture that it keeps the same form for any PM order.
If we take the reduced energies as $\hat{\mathcal{E}_0}=\mathcal{E}_0/m_0$ and $\hat{\mathcal{E}}=\mathcal{E}/m_0$, the energy map can be recast as
$\hat{\mathcal{E}_0}=\frac{\nu}{2}\hat{\mathcal{E}}^2-\frac{1}{2 \nu}+1$, which is useful in the related calculations.

The parameters ($a_i, \ d_i$) for $i=2,~3,~4$ appearing in the effective metric \eqref{Mmetric} are found by identifying the scattering angles for the two systems order by order, i.e., by taking $\chi^{\text{real}}_i=\chi^{\text{eff}}_i\;, $ for $i=2,~3,~4$.
However, we note that one  equation corresponds to two unknowns $a_i$ and $d_i$ for each order $i$. Thus, one of them, say $d_i$, can be considered as an undetermined parameter which should be fixed in the specific application. It should be noted that the effective metrics at the 3PM and 4PM orders, described by  Eq. \eqref{Mmetric}  in the Schwarzschild-like coordinates, include the  terms $\chi^{rr}_3$ and $\chi^{rr}_4$, which  represent the 3PM and 4PM radiation-reaction effects, respectively. In other words, the structure of the effective spacetime will be affected by the radiation-reaction effect.

For the general effective metric \eqref{Mmetric} with the parameters \eqref{parameters}, there are three  non-vanishing null tetrad components of the tracefree Ricci tensor: $\phi_{00},$ $ \phi_{11} $ and $\phi_{22}$. We cannot isolate the decoupled equation for the null tetrad component of gravitationally perturbed Weyl tensor $\Psi_4^B$ in the effective spacetime because too many Newman-Penrose equations are coupled with each other.  However, if we take $ D=1$ (i.e., $d_i=0$ for $i=2,3,4$) which fixed the freedoms of the parameters of the effective metric, we have $\phi_{00}=\phi_{22}=0 $ and we can obtain the decoupled equation for the null tetrad component of  the gravitationally perturbed Weyl tensor $\Psi_4^B$ by means of the Refs.  \cite{Jing,Jing1,Jing2} which showed that the decoupled equation for $\Psi_4^B$ can be obtained for the spacetime with the parameter $D=1$ up to any PM order. Then, based on the results of this paper, we can set up  a  self-consistent EOB theory in which the Hamiltonian, radiation-reaction forces  and waveforms for the ``plus" and ``cross" modes of the gravitational wave are based on the same effective spacetime.

\vspace{0.3cm}
\acknowledgments
{\it Acknowledgments}: We would like to thank professors S. Chen and Q. Pan  for useful discussions on the manuscript. This work was supported by the Grant of NSFC Nos. 12035005 and 12122504, and National Key Research and Development  Program of China No. 2020YFC2201400.  

\bibliography{refs}

\end{document}